# Role of Thermal Resistance on the Performance of Superconducting Radio Frequency Cavities

Pashupati Dhakal, Gianluigi Ciovati and Ganapati Rao Myneni
Jefferson Lab, Newport News, VA 23606, USA

E-mail: dhakal@jlab.org

**Abstract**
   Thermal stability is an important parameter for the operation of the superconducting radio frequency (SRF) cavities used in particle accelerators. The rf power dissipated on the inner surface of the cavities is conducted to the helium bath cooling the outer cavity surface and the equilibrium temperature of the inner surface depends on the thermal resistance. In this manuscript, we present the results of direct measurements of thermal resistance on 1.3 GHz single cell SRF cavities made from high purity large grain and fine grain niobium as well as their rf performance for different treatments applied to outer cavity surface in order to investigate the role of the Kapitza resistance to the overall thermal resistance and to the SRF cavity performance. The results show no significant impact of the thermal resistance to the SRF cavity performance after chemical polishing, mechanical polishing or anodization of the outer cavity surface. Temperature maps taken during the rf test show non-uniform heating of the surface at medium rf fields. Calculations of $Q_0(B_p)$ curves using the thermal feedback model show good agreement with experimental data at 2 K and 1.8 K when a pair-braking term is included in the calculation of the BCS surface resistance. These results indicate local intrinsic non-linearities of the surface resistance, rather than purely thermal effects, to be the main cause for the observed field dependence of $Q_0(B_p)$.

## 1. Introduction

Superconducting radio frequency cavities are the building blocks of particle accelerators for basic physics research. They are based on niobium superconducting hollow structures ("cavities") to accelerate the beam of charged particles. The superiority of the superconducting material is its ability to efficiently store large amount of energy with no or very little dissipation. The performance of SRF cavities is measured in terms of the quality factor expressed as $Q_0 = \omega U/P_{diss}$, where $U$ is stored energy and $P_{diss}/\omega$ is the power dissipation on the inner cavity wall per rf cycle. Ideally, the quality factor of SRF cavities is independent of the accelerating field (or peak magnetic field) as the breakdown occurs at the superheating field. However, due to the finite resistance of the superconductor in an rf field, power dissipation occurs on the inner cavity walls due to the interaction of rf field with normal conducting electrons. At increasingly higher peak surface magnetic field, the surface resistance is expected to increase due to pair-breaking by strong rf field giving rise to a non-linear BCS surface resistance [1]. The highest rf field achieved in cavities is often limited by a local quench of the superconducting state due to thermal, magnetic, or magneto-thermal instabilities at "weak" superconducting regions or at normal-conducting defects [2]. The dissipated power density is given by $P_{diss} = \frac{1}{2} R_s(T) H^2$, where $R_s(T)$ is surface resistance, and $H$ is local rf magnetic field on the inner surface of the cavity. The power dissipated (heat) on the inner surface of SRF cavities during operation is conducted through the cavity wall into the helium bath. The efficient transport of heat from the inner cavity wall to the helium bath depends on the thermal conductivity of niobium and the Kapitza conductance between the outer cavity surface and the superfluid helium. The thermal conductivity of niobium is related to the residual resistivity ratio, RRR, and it is material dependent, for example on the impurities content, crystal grains size, defects and dislocations in niobium. The Kapitza resistance is an intrinsic thermal resistance due to the phonon mismatch at the boundary between niobium and the superfluid helium and depends on the nature of the solid surface, such as the presence of oxides, foreign materials and roughness.

The role of thermal resistance on the performance of SRF cavities hasn't been fully understood. Most of the SRF cavities performance is limited due to phenomena such as high field $Q$-slope, field emission, multipacting, and quench. The origin of these non-linear power dissipation mechanisms at high accelerating field is still an open area of research. Besides the causes for the degradation of the quality factor at high rf field, $B_p \gtrsim 90$ mT, it is important to understand the causes for the reduction of $Q_0$ in the medium field range (20-90 mT), referred in the literature as "medium-field $Q$-slope", as continuous-wave SRF accelerators rely on moderate gradients but high quality factors for efficient operation.

Poor heat transfer from the inner cavity surface to the helium bath can affect the cavity performance by both reducing $Q_0$ with increasing rf field and lowering the quench field. As explained by the so-called thermal feedback model (TFBM) [1], the thermal boundary resistance provides the positive feedback mechanism to the temperature gradient between the inner cavity surface and the He bath with increasing rf field, which is then amplified by the BCS surface resistance, through its exponential temperature dependence, up to the point of thermal instability, triggering a quench. In the case of normal-conducting defects, poor heat transfer causes the local temperature at the defect to increase rapidly with increasing rf field, causing a quench when the local temperature exceeds the critical temperature ($T_c$) of the superconductor surrounding the defect. Mitigation of quenches due to such normal defects was the main reason to push for high thermal conductivity (high RRR) Nb to fabricate SRF cavities [2]. Calculations of the $Q_0(B_p)$ curves using the thermal feedback model have been reported in several articles [3,4]. Comparisons with experimental data showed that a good agreement could be found, in most cases, when a pair-breaking term is added to the standard BCS surface resistance used to calculate the power dissipation with the TFBM [1]. Values of thermal conductivity $\kappa$, and Kapitza resistance $R_K$ of Nb taken from the literature on Nb samples are used in the calculation of $Q_0(B_p)$. However, there could be a significant uncertainty on such values as they depend strongly on the phonon mean free path in the Nb and the conditions of the outer cavity surface.

In this contribution, we present the result of thermal resistance measurements directly on SRF cavities as well as their rf performances. The thermal feedback model is then applied, using the measured thermal resistance, to compare the calculated $Q_0(B_p)$ curves with the experimental ones. Furthermore, we have applied the temperature mapping technique to map the temperature of the SRF cavity surface during rf tests at or below 2.0 K to identify the hot spots and quench locations in order to distinguish between uniform and localized heating.

## 2. Experimental Setup

Two 1.3 GHz single-cell cavities of the TESLA/XFEL shape [5], one made from large grain Nb from Tokyo Denkai with RRR> 250 (labeled TD5) and other made from fine grain from Ningxia with RRR >250 (labeled RDT13) were used in this study.

### A. Thermal resistance measurement

Previously, thermal resistance measurements on SRF niobium samples were carried out in an experimental cell and supplemental measurements of the thermal conductivity allowed extracting Kapitza resistance [6,7,8]. Palmieri *et al.,* [9] measured the thermal boundary resistance via the rf surface resistance measurement in SRF cavities and reported a decrease in thermal boundary resistance after anodizing the outer surface of the cavity. In our present study, we have estimated the thermal resistance of SRF cavity using the method similar to that used to characterize Nb samples in refs. [6,7]. The schematic representation of the experimental set up is shown in figure 1.

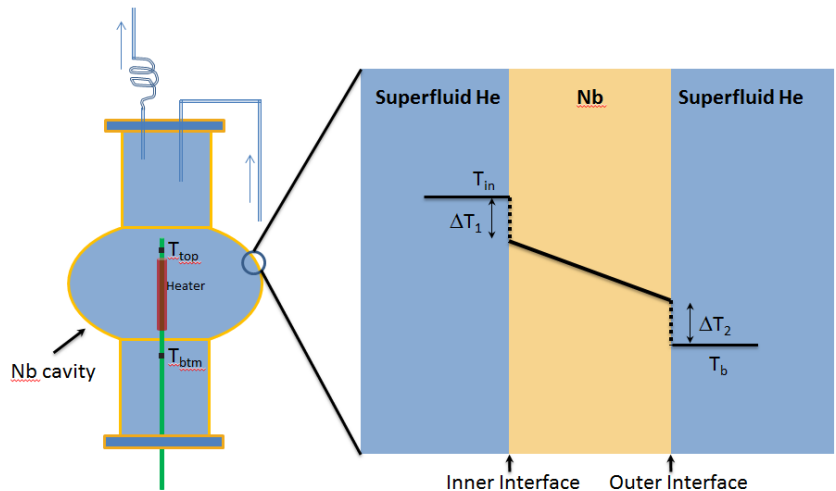

**Figure 1.** Schematic of experimental set up to measure the thermal resistance.

The niobium SRF cavity of thickness ~ 2.9 mm is immersed in superfluid helium bath and filled with superfluid helium via a capillary tube of diameter ~1.5 mm. Two capillary tubes are used, one for filling and the other for exhausting the He gas during filling. The capillary tubes are welded to a ~9.5 mm thick stainless steel flange sealed to one cavity flange with indium wire. A cryogenic heater of resistance ~8.5 Ω is inserted along the axis of the cavity using a G10 rod. Two calibrated Cernox thermometers are attached at the both ends of the G10 rod, measuring the temperature at the middle of cavity and at the beam tube. The outside temperature is regulated via He pumping and the temperature is held to within ±1 mK. In the absence of any additional heat source, the superfluid He inside the cavity is in thermal equilibrium with the bath temperature. The temperature rise inside the cavity is measured as a function of the applied power as shown in figure 2(a). Due to the high thermal conductivity of the superfluid helium, thermal equilibrium is achieved quickly and the power density on the inner cavity surface is assumed to be uniform. The surface area of the single cell cavity with beam tubes is ~1794 $cm^2$. The area of the stainless steel blank flanges is ~153 $cm^2$. Given the thermal conductivity of stainless steel at 2 K being ~0.1 W/m K [10], the total heat loss through the flanges is estimated to be less than 2 % of the total heat loss. The critical heat flux through the capillary tube varies with temperature from 1.5 K to the lambda point, between ~80-180 mW/$cm^2$, with a maximum at ~1.7 K [11]. In our experimental setup, the maximum heat loss via superfluid helium inside the cavity is estimated to be less than 2%. The heat loss via the electrical feedthrough is negligible. Thus ~96 % of heat was carried away across the cavity. Beam tubes were made from low purity reactor grade niobium (RRR~60) and having thermal conductivity at 2 K of ~1 W/m K [12]. The results presented in this contribution are the combined effect for both beam-tubes and cavity.

Under the steady state condition with power density $q$ dissipated at the inner surface of the cavity, the temperature jump $\Delta T = T_{in} - T_b$ is established between the He bath at temperature $T_b$ and the inner cavity volume at temperature $T_{in}$. The temperature difference can be written as $\Delta T = T_{in} - T_b = R_B q$, where $R_B$ is the thermal resistance. Therefore, the slope in the plot $\Delta T$ vs $q$ (shown for example in figure 2 b) yields $R_B = d/\kappa + R_K$, where $d$ is the thickness of the wall, $\kappa$ is the thermal conductivity, $R_K$ represents the thermal resistance between the cavity wall and the superfluid He (Kapitza resistance). The measured thermal resistance is the sum of the contribution from the stainless steel and niobium cavity. Since ~96% of heat is carried away across the cavity, the contribution to the thermal resistance due to stainless steel can be neglected. It should be noted that since there are two interfaces (inner and outer surface) between niobium and superfluid He, $R_K = R_{K,in} + R_{K,out}$. The unit of the thermal resistance will be $cm^2$ K/W throughout this article. Different treatments were applied to the outer cavity surfaces in order to investigate the effect of surface preparations on the thermal resistance.

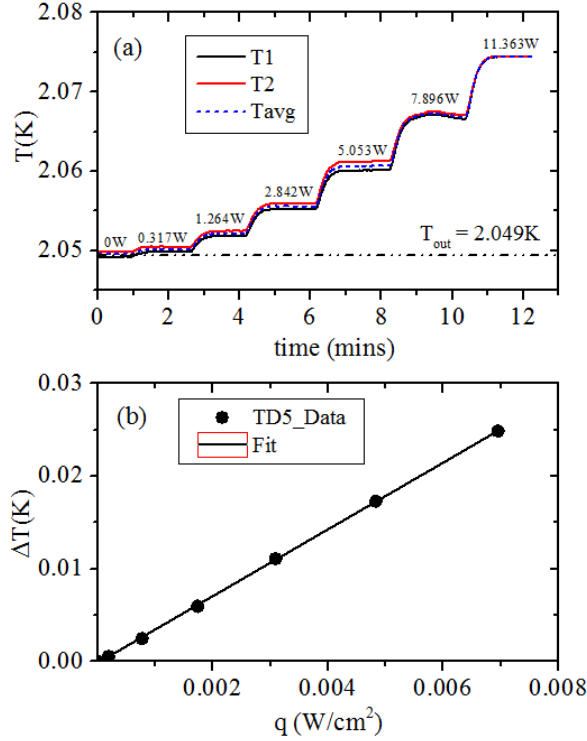

**Figure 2.** Typical experimental data measured. (a) The increase in temperature of superfluid He inside the cavity as a function of applied power at 2.05 K and (b) the plot of $\Delta T$ vs $q$ for cavity TD5. The slope of the fit gives the thermal resistance at 2.05 K.

### B. RF Test and temperature mapping

Standard cavity preparation procedures were adopted before the cavity tests [13]. The rf measurement consisted of measuring $Q_0$ vs $T$ at low peak surface magnetic field $B_p \sim 10$ mT between 4.2-1.6 K. The surface resistance was calculated as $R_s = G/Q_0$, where $G$ is the geometry factor, and $R_s(T)$ was fitted with $R_s(T) = R_{BCS}(T)+R_{res}$ to extract the superconducting gap, and the residual resistance. $Q_0$ vs $B_p$ data were taken at 2.0, 1.8 and 1.6 K to the highest field.

To further understand the effect of the thermal resistance on the performance of SRF cavity, the large grain cavity TD5 was measured with a temperature mapping system built at Jefferson lab [14], based on the system developed at Cornell University [15]. The temperature mapping system consists of the custom-made sensors from 100 Ω Allen-Bradley carbon resistors (5%, 1/8 W) calibrated against a calibrated Cernox temperature sensor immersed in the helium bath. In our current experiment, 540 sensors cover the cavity surface with 15 sensors on each vertical board, spaced azimuthally 10° apart. Each sensor on the board is labeled from 1 to 15, with the sensor 1 closer to the top iris of the cavity, sensor 8 at the equator, and sensor 15 close to bottom iris. The temperature maps were taken during the high power rf tests at 2.0, 1.8 and 1.6 K.

## 3. Experimental Results

### A. Thermal resistance

The single-cell cavities used for this study were previously doped with nitrogen and earlier rf measurements were carried out and were published in ref. [16]. The cavities were heat treated at 800 °C for 3 hours followed by 20 minutes of exposure to nitrogen at 25 mTorr at this temperature. The inner surface of cavity TD5 was electropolished (EP) to remove ~40 μm from the inner surface and ~20 μm was subsequently removed from the outer surface by buffered chemical polishing (BCP). The results of measurements after inner EP and outer BCP [17] showed no significant change in thermal resistance even though the cavity's rf performances are significantly different [16]. The baseline test (test 1) in figure 3 refers to the data taken after N-doping, inner EP and outer BCP. The outer surface of the cavity TD5 was modified by surface roughening using sand paper (100μm), anodization (~50nm thick oxide) at 25 V with ammonium hydroxide [18] and additional BCP (~2 μm) to remove the oxide layer. The results from thermal resistance measurements are shown in figure 3.

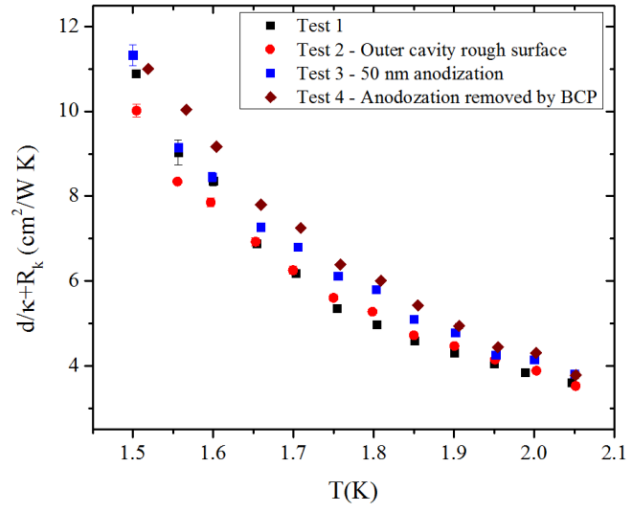

**Figure 3**. Results of the thermal resistance measurement on ingot Nb cavity TD5.

The thermal resistance of cavity RDT13 measured after the removal of the N-doping (~ 40μm inner surface by EP and ~20 μm by BCP) is labeled as baseline (test 1). After the baseline measurement the outer surface of the cavity was subjected to ~50 nm anodic oxidation and etching of the outer surface by BCP (~2 μm) followed by the low temperature baking at 120 °C for 48 hours. After each surface modification, rf tests were also performed on both cavities and the results will be presented in the next section. The measured thermal resistance on the fine-grain cavity is shown in figure 4.

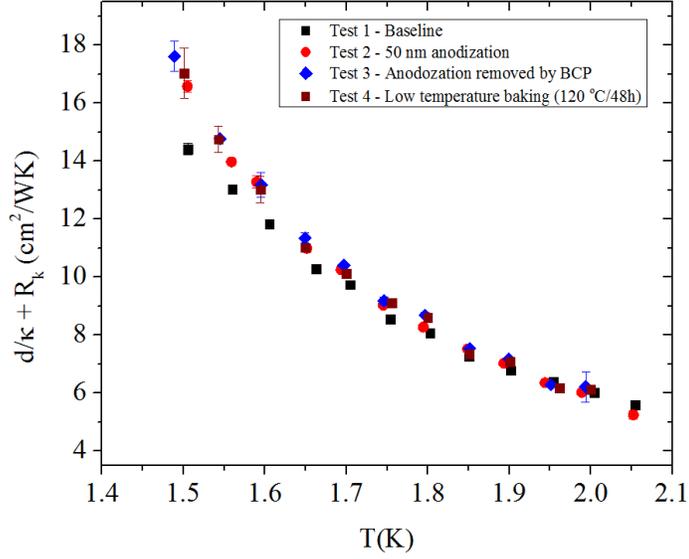

**Figure 4.** Results of the thermal resistance measurement on fine-grain cavity RDT13.

## B. RF test

The RF test was conducted by exciting the cavity in TM$_{010}$ mode using the phase-locked loop RF system to measure the incident, reflected and transmitted powers along with the resonant frequency and decay time to calculate the quality factor and accelerating gradient/peak magnetic field [19]. The surface resistance averaged over the whole cavity surface was obtained as $R_s = G/Q_0$ from $Q_0$ measured at low peak magnetic field, $B_p \sim 10$ mT from 4.2-1.6 K and fitted with $R_s(T_{in}) = R_{BCS}(T_{in}) + R_{res}$ where $R_{BCS}(T_{in})$ is given by a common approximation of the surface resistance calculated from the BCS theory, valid at $T \ll T_c$ and in the limit of zero rf field:

$$R_{BCS}(T_{in}) = \frac{A}{T_{in}} e^{-U/k_B T_{in}}, \quad (1)$$

where $A$ is a factor related to material parameters, such as the penetration depth, coherence length and mean free path, and frequency, $U$ represents the energy gap and $k_B$ is Boltzmann's constant. Since the measurements were taken at very low field, overheating of the inner surface can be neglected and $T_{in}$ was taken to be the same as the measured He bath temperature. The parameters extracted from the $R_s(T)$ curves are presented in Table 1.

**Table 1**. Fit parameters extracted from fitting $R_s(T)$ data for cavities TD5 and RDT13 with Eq. (1)

| Cavity ID | Outer Surface Preparation | $A$ ($10^{-4}\,\Omega$ K) | $U$ (meV) | $R_{res}$ (n$\Omega$) |
|---|---|---|---|---|
| TD5 | Baseline | 2.06±0.01 | 1.57±0.01 | 2.8±0.1 |
| | +Sanding | 2.09±0.02 | 1.54±0.01 | 3.0±0.2 |
| | +50 nm anodization | 2.03±0.02 | 1.52±0.01 | 2.7±0.2 |
| | +BCP | 2.16±0.02 | 1.53±0.02 | 2.5±0.1 |
| RDT13 | Baseline | 2.14±0.01 | 1.53±0.01 | 2.8±0.1 |
| | +50 nm anodization | 2.01±0.02 | 1.55±0.01 | 3.0±0.3 |
| | +BCP | 1.96±0.01 | 1.54±0.01 | 2.8±0.3 |
| | +LTB (120 °C/48hrs) | 1.42±0.03 | 1.53±0.02 | 6.7±0.1 |

Measurements of $Q_0(B_p)$ were done at 2.0, 1.8 and 1.6 K up to the breakdown field, $B_{p,max}$. The typical experimental uncertainties are ~10% and ~5% for $Q_0$ and $B_p$ measurement, respectively. Figure 5 shows the $Q_0$ vs $B_p$ data for cavity TD5 at 2 K and 1.6 K after different outer surface treatments. All rf tests were limited by quench. In tests 1 and 4, $Q_0$ vs $B_p$ data for cavity TD5 showed some multipacting starting at $B_p$~ 75 mT, which resulted in a slight drop of $Q_0$. There is no significant change in quality factor as a result of outer surface modifications. There was no field emission in any of the tests.

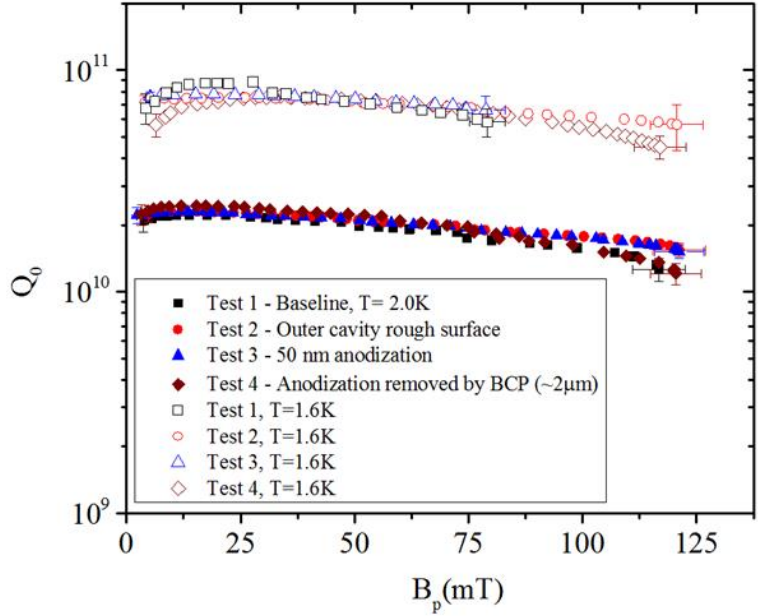

**Figure 5**. $Q_0$ vs $B_p$ data for ingot Nb cavity TD5 at 2.0 K (solid) and 1.6K (empty) after different outer surface modification. The rf measurement during tests 1 and 3 at 1.6 K were stopped at $B_p$ ~ 75 in order to remain below the multipacting barrier.

$Q_0$ vs $B_p$ data for cavity RDT13 at 2 K and 1.6 K, shown in figure 6, showed no significant change in the $Q_0$-values between $B_p$ ~20-110 mT as a result of outer surface modifications. The cavity has a low field $Q$-rise $B_p < 20$ mT, medium field $Q$-slope $20 < B_p < 110$ mT, and high field $Q$-slope $B_p > 110$ mT. The rf tests 1-3 were limited by the high-field Q-slope. Test 4 was limited by quench at $B_p$ ~ 153 mT. There was no field emission in any of the tests.

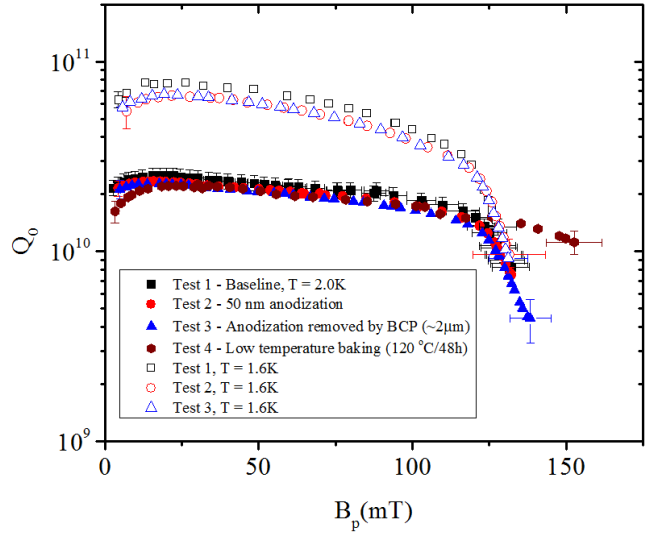

**Figure 6.** $Q_0$ vs $B_p$ data for fine-grain cavity RDT13 at 2.0 K (solid symbol) and at 1.6 K (empty symbols) after different treatments applied to the outer surface.

### C. RF test with temperature map

As mentioned earlier, rf measurements on cavity TD5 were repeated with temperature mapping after test 4. Temperature maps were acquired at 2.0, 1.8 and 1.6 K while increasing the rf field below the quench value. Figure 7 shows maps of the temperature of the outer surface relative to that of the He bath, $\Delta T_{out}$, taken at 1.6 K at ~62 mT and at ~116 mT. Figure 8 (a) shows a temperature map at 2 K and $B_p$ ~ 116 mT just before quench and (b) shows $\Delta T_{out}$ vs $B_p$ at two different locations during the RF test at 2.0 K. The temperature maps clearly show that weak overheating ($\Delta T_{out}$ ~ 10 mK) develops at several spots of the cavity surface with increasing rf field. A fit of $\Delta T_{out} \propto B_p^m$ for sensors 40-8 and 120-7 shown in fig. 8(b) resulted in $m$ ~ 2.5 suggesting that the local power dissipation is stronger than simple Joule heating. Sensor location 320-7, showed relatively small heating with m ~ 2.

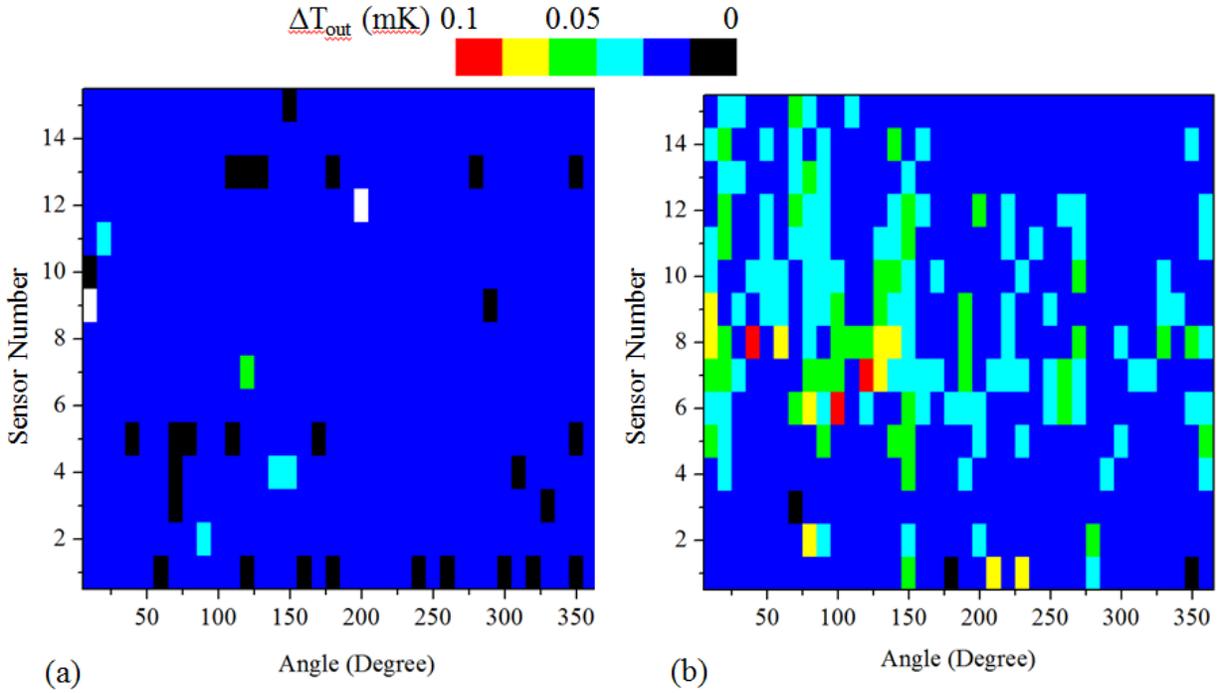

**Figure 7.** "Unfolded" temperature maps at 1.6 K on cavity TD5 at (a) $B_p \sim 62$ mT and (b) at $B_p \sim 116$ mT.

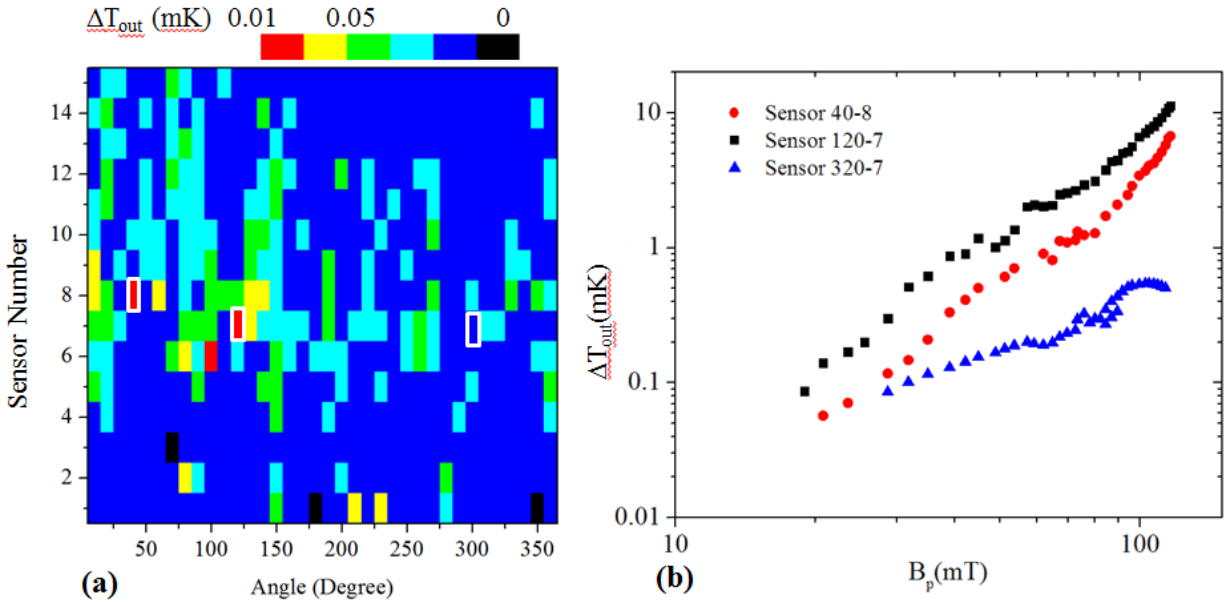

**Figure 8.** "Unfolded" temperature map at 2.0 K on cavity TD5 (a) just before quench ($B_p \sim 116$ mT) and (b) $\Delta T_{out}$ measured at three different locations highlighted with a white border in (a), during increasing rf power.

During the final rf test at 1.8 K, the field in the cavity was increased until it quenched at ~120 mT and a snapshot of the temperature map during quench is shown in figure 9. The quench location showed no significant precursor heating prior to quench, suggesting the possibility of magnetic, rather than thermal, origin of quench.

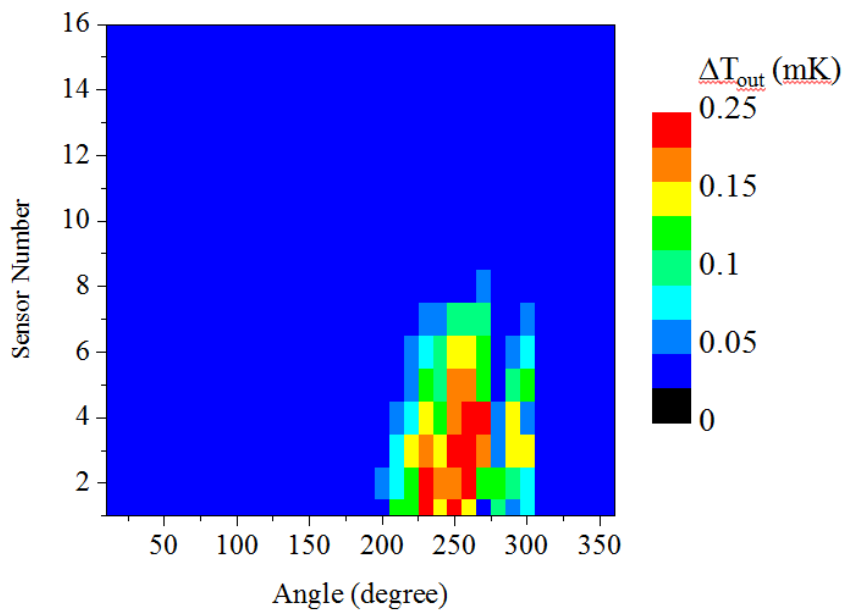

**Figure 9.** Temperature map taken during quench at 1.8K at $B_p \sim 120$ mT.

## 4. Analysis of Experimental Data

### A. Thermal resistance

The experimental results show modest (10-15% variation) effect of outer surface preparations on the thermal resistance. Earlier measurements on cavity TD5 doped with nitrogen also showed little effect of outside BCP and inside EP on the thermal resistance, even though the cavity's rf performances were significantly different [16, 17]. The simplest way to describe the temperature dependence of the thermal resistance is with a power law $R_B = a\,T^n$, being related to the power law dependence of $R_K(T) \propto 1/T^3$ [20] and of $1/\kappa(T) = \rho_0/L_0 T + bT^2$. $\rho_0$ is the residual resistivity, $L_0$ is the Lorentz constant and $b$ is the coefficient of momentum exchange with the lattice vibrations. The fit coefficients $a$ and $n$ for both cavities, after different treatments, are listed in Table 2. The measurement on RDT13 cavity showed no significant change on thermal resistance by different outer surface treatments. It is known that low temperature baking reduces the high field $Q$-slope in SRF cavities, which is also shown in fig. 6. The data presented in Sec. 3A showed no significant change of the thermal resistance before and after baking. As the dissipated power shown in fig. 6 is nearly identical at 110 mT but significantly different at higher field before and after baking, it implies that the improvement of the quality factor resulting from the 120 °C bake cannot be explained by a reduced thermal resistance.

The measured thermal resistance on RDT13 in the temperature range 2.1-1.5 K is about 50% higher than the one measured for TD5 cavity although they have similar temperature dependence, $T^n$, $n \sim 3$–4. Assuming the Kapitza conductance to be the same in both cavities, the thermal conductivity in the fine grain cavity would need to be about a factor of three lower than that of the large grain cavity in this temperature range to explain the difference in thermal resistance. Higher thermal conductivity below ~2.1 K in large grain niobium compared to fine grain one has been measured in the presence of the so-called phonon peak [21,22].

Table 2. The fit coefficients from the power law $R_B = aT^n$.

| Cavity ID | Outer Surface Preparation | $a$ (cm$^2$ W/K$^{1-n}$) | $n$ |
|---|---|---|---|
| TD5 | Baseline | 51±1 | 3.8±0.1 |
| | +Sanding | 31±1 | 3.2±0.1 |
| | +50 nm anodization | 39±1 | 3.3±0.1 |
| | +BCP | 47±2 | 3.6±0.1 |
| RDT13 | Baseline | 50±2 | 3.1±0.1 |
| | +50 nm anodization | 69±2 | 3.6±0.1 |
| | +BCP | 70±2 | 3.6±0.1 |
| | +LTB (120 °C/48hrs) | 50±1 | 3.5±0.1 |

### B. Analysis of rf data with thermal resistance

The measurements of the cavities' low-field surface resistance and of the thermal resistance allow us to calculate the dependence of $Q_0$ on $B_p$ on the basis of the TFBM. This is done by calculating $T_{in}(B_p)$ by solving the heat balance equation given by

$$\frac{1}{2} R_s(T_{in}, H_p) H_p^2 = \frac{(T_{in} - T_b)}{R_B} \qquad (2).$$

In the absence of any intrinsic field dependence of the surface resistance, $R_s(T_{in})$ in eq. (2) is given by the sum of the BCS surface resistance, given by eq. (1), and $R_{res}$. The calculated quality factor along with the rf data is shown in figure 10 for cavities TD5 and RDT13 and shows that the TFBM significantly underestimates the decrease of $Q_0$ with increasing rf field at all temperatures. Gurevich estimated that the pair-breaking effect caused by an increasing rf field would result in an intrinsic field dependence of the BCS surface resistance which can be approximated in the clean limit by the following eq. (1):

$$R_{BCS}(H_p) \cong \left[1 + \gamma(T_b) \left(\frac{H_p}{H_c}\right)^2\right] R_{BCS,0} \qquad (3)$$

where $R_{BCS,0}$ is the linear BCS resistance given by eq. (1), $H_c$ is the superconducting critical field and $\gamma(T_b) = \frac{\pi^2}{384}\left(\frac{U}{k_B T_b}\right)^2$. Figure 10 also shows the field dependence of the quality factor calculated from eq. (1) in which the non-linear BCS surface resistance given by eq. (3) was used. $H_c = 200$ mT was used in the calculation. There is good agreement between the data and the model at 2.0 K up to the onset of the high-field $Q$-slope, observed in RDT13, whereas the model underestimates the decrease of $Q_0$ with increasing field at 1.6 K.

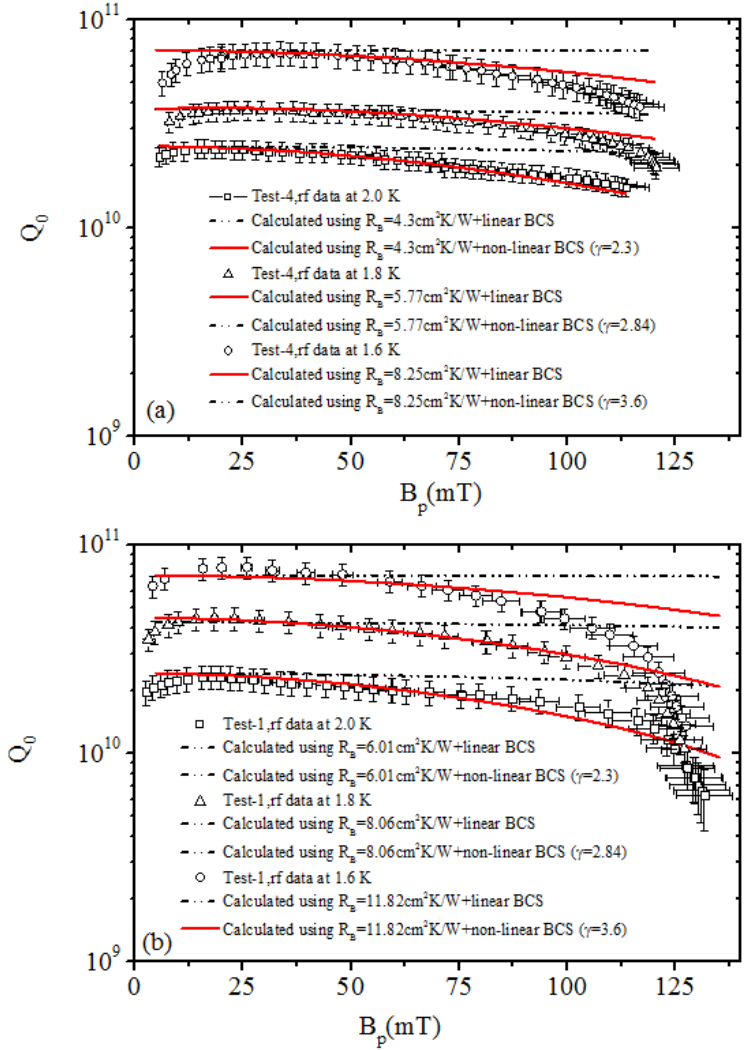

**Figure 10.** Comparison between measured and calculated $Q_0(B_p)$ curves for cavity TD5 (a) and RDT13 (b) using the TFBM with the measured thermal resistance for the cases of linear (--) and non-linear BCS resistance (—) as described in the text.

## 5. Discussion

The results from direct measurements of the thermal resistance of the SRF Nb cavities presented in this article showed a weak dependence from treatments such as sand blasting, BCP and anodization applied to the outer surface. The temperature dependence and the values of thermal resistance we have measured are consistent with the results of Kapitza resistance measurements on small, flat samples [6,7,8] and of thermal resistance obtained from rf measurement of cavities [9]. The small variations of thermal resistance with outer surface treatments can be related to the thermal conductivity of the Nb being the dominant term. For example, it was shown in [6] that the Kapitza resistance would become the dominant term of the thermal resistance only after a post-purification annealing which strongly reduces κ.

A decrease of thermal resistance by a factor of ~3 by anodization of the outer surface was recently reported. The thermal resistance was extracted from measurements of the surface resistance as a function of temperature at constant dissipated power in 6 GHz single-cell cavities. The results from the high-power tests of the 6 GHz cavity also showed higher $Q_0$ and quench field values after anodization of the outer surface. However, those results also showed a higher $Q_0$ at very low field and nearly identical medium field Q-slope after anodization, which are difficult to understand solely on the basis of reduced thermal resistance. The results we obtained in 1.3 GHz cavities showed no significant variation of either thermal resistance or $Q_0(B_p)$ by anodization.

Calculations of $Q_0(B_p)$ curves with the TFBM using the linear BCS surface resistance shown in Fig. 10 clearly show that:

- the model significantly underestimates the measured decrease of $Q_0$ with increasing rf field.
- if an intrinsic field dependence of the surface resistance, such as that due to a pair-breaking term, is added to the model then there is very good agreement between the calculation and the experimental data at 2 K and 1.8 K, without introducing any fit parameter. However, increasing discrepancies occur at lower temperatures.

Such conclusions are consistent with the results from a similar analysis applied to a broad range of cavity test data at different temperatures and frequencies, published in Ref. 3.

The temperature maps reveal that there is a large fraction of the cavity surface which shows very weak overheating, consistent with what predicted by the TFBM with linear $R_{BCS}$, however there are many regions with enhanced dissipation. In those regions, the dissipation is stronger than Joule-type heating and it causes the $Q_0$ to decrease more strongly with increasing rf field. The non-uniformity of the surface resistance in SRF bulk Nb cavities was also demonstrated by using the laser scanning microscopy technique [23]. Different dissipation mechanisms in different regions of the cavity can be related to the local distribution of interstitial impurities, oxides stoichiometry, precipitates or to the presence of pinned vortices.

Recent advances in the processing of bulk niobium cavities, such as controlled doping of Nb with titanium [24] or nitrogen [25,26] interstitial impurities within the rf penetration depth, resulted in remarkable cavity performances in which the quality factor increases by up to a factor of ~2 with increasing field up to ~80 mT. Of course, such $Q_0(B_p)$ dependence would be completely unaccounted for by the TFBM or improvements in the thermal resistance. The *Q*-rise phenomenon has been recently explained in terms of an intrinsic feature of the BCS surface resistance in the non-equilibrium dirty limit [27].

The above considerations and the experimental results shown in Sec. 3 lead us to the conclusion that the thermal resistance has a very limited influence on determining the field dependence of the surface resistance of bulk Nb cavities in the GHz range and at temperatures below the lambda point. Rather, we have provided evidence for the dependence of $Q_0(B_p)$ being dictated by local intrinsic non-linearities of the surface resistance which can result in a quality factor which increases, decrease or independent from the rf field and there exist a lot of data in the literature in which each of these three dependencies have been measured.

The role of the thermal boundary resistance certainly becomes significant in bulk Nb cavities above the lambda point, because of the poorer heat transfer properties of He I, as it has been noted in [9, 28, 29]. The role of the thermal boundary resistance can also be significant in thin film cavities, if the film has regions with poor adhesion to the substrate, as discussed in Ref. [30], of if the film has low thermal conductivity, such as $Nb_3Sn$.

## 6. Conclusion

We have measured the thermal resistance of fine-grain and large-grain SRF niobium cavities subjected to several surface preparations. The results showed no significant variation of the thermal resistance by BCP, anodization or mechanical polishing of the outer surface, which suggests that thermal conduction through the cavity walls dominates the thermal resistance of the cavities we have tested. High-power rf test of the cavities at 2 K, 1.8 K and 1.6 K with temperature mapping revealed regions with different dissipation mechanisms at medium rf fields. Comparisons of measured $Q_0(B_p)$ curves with those calculated from the TFBM, without any fit parameter, showed a good agreement at 2 K and 1.8 K when including an intrinsic pair-breaking term in the BCS surface resistance, the agreement becoming poorer at lower temperature. These results, along with those published in the literature, lead us to the conclusion that the thermal resistance has a limited influence on the field dependence of the surface resistance of bulk Nb cavities at or below 2 K and that such dependence is rather driven by local intrinsic non-linearities of the surface resistance which are likely being related to the distribution of impurities within the rf penetration depth.

## 7. Acknowledgments

We would like to acknowledge Jefferson Lab SRF technical staffs for help with the cavity annealing, HPR, EP and cryogenic operations. We would like to acknowledge A. Palczewski and P. Kneisel for providing cavities RDT13 and TD5 for this study. This manuscript has been authored by Jefferson Science Associates, LLC under U.S. DOE Contract No. DE-AC05-06OR23177. The U.S. Government retains a non-exclusive, paid-up, irrevocable, world-wide license to publish or reproduce this manuscript for U.S. Government purposes.